# Overcoming Anchoring Bias: The Potential of AI and XAI-based Decision Support

*Completed Research Paper*

**Felix Haag**
University of Bamberg
An der Weberei 5,
96049 Bamberg, Germany
felix.haag@uni-bamberg.de

**Carlo Stingl**
University of Bamberg
An der Weberei 5,
96049 Bamberg, Germany
carlo.stingl@uni-bamberg.de

**Katrin Zerfass**
University of Bamberg
An der Weberei 5,
96049 Bamberg, Germany
katrin.zerfass@gmail.com

**Konstantin Hopf**
University of Bamberg
An der Weberei 5,
96049 Bamberg, Germany
konstantin.hopf@uni-bamberg.de

**Thorsten Staake**
University of Bamberg
An der Weberei 5,
96049 Bamberg, Germany
thorsten.staake@uni-bamberg.de

## Abstract

*Information systems (IS) are frequently designed to leverage the negative effect of anchoring bias to influence individuals' decision-making (e.g., by manipulating purchase decisions). Recent advances in Artificial Intelligence (AI) and the explanations of its decisions through explainable AI (XAI) have opened new opportunities for mitigating biased decisions. So far, the potential of these technological advances to overcome anchoring bias remains widely unclear. To this end, we conducted two online experiments with a total of N=390 participants in the context of purchase decisions to examine the impact of AI and XAI-based decision support on anchoring bias. Our results show that AI alone and its combination with XAI help to mitigate the negative effect of anchoring bias. Ultimately, our findings have implications for the design of AI and XAI-based decision support and IS to overcome cognitive biases.*

## Introduction

Decisions are omnipresent in peoples' everyday lives and differ by several characteristics, such as the degree of complexity and the cognitive effort required. In particular, the latter is an essential parameter for the decision-making process: Human decisions often rely on intuition or gut feeling (Tversky & Kahneman, 1974). Being aware of this decision mode, behavioral theorists have frequently found human decision-makers to be subject to limited information processing capabilities and bounded rationality (Simon, 1990).

Consequently, instead of deciding entirely rationally, humans are prone to heuristics and biases in decision-making (Goodwin & Wright, 2014; Kahneman, 2011).

Among many researched heuristics and biases, the anchoring bias is one of the most robust phenomena in human decision-making (Furnham & Boo, 2011). The anchoring bias occurs when an initial given value (i.e., the anchor) affects an individual's subsequent decision—even if the anchor is irrelevant, arbitrary, or uninformative for the actual decision task (Kahneman, 2011). While initially investigated as a phenomenon that emerges for numerical estimation tasks such as multiplication exercises (Tversky & Kahneman, 1974), the existence of anchoring bias has already been demonstrated for various kinds of decision-making processes, e.g., the formation of economic preferences in the context of purchases (Ariely et al., 2003). The anchoring bias can impose severe consequences for decision-makers (Kahneman, 2011), especially when used with deliberately negative intent, which is also referred to as "dark pattern" (Sin et al., 2022). Examples of these manipulations can frequently be found in e-commerce, such as using pre-discounted products as anchors (e.g., higher strikethrough prices than the actual price) to trigger impulsive purchases (Moser, 2020) or expanding an individual's purchase quantity over initially held levels (Simonson & Drolet, 2004; Wansink et al., 1998). As this effect also holds for interactions in the digital space, information systems (IS) research has long been investigating ways to effectively support decision-makers to reduce anchoring bias, e.g., by displaying warning messages alongside positioned anchors (George et al., 2000).

One promising field that has recently gained traction to support decision-makers in overcoming heuristic biases is Artificial Intelligence (AI) (Wang et al., 2019). AI employing complex machine learning (ML) models has the capability to detect patterns in data and make precise predictions, which can result in more informed decisions and higher task performance among individuals (Fügener et al., 2021). The application of AI has recently been shown to be effective in mitigating anchoring bias by predicting anchored decisions and reframing decision environments without making individuals explicitly aware (Echterhoff et al., 2022). However, predictions that are usually obtained from inscrutable "black box" ML models are only one source of information for decision support—in particular, explaining the patterns that complex models discover in the data promises significant benefits for human decision-making (Adadi & Berrada, 2018). This has fueled the development of explainable AI (XAI) methods that translate the patterns of complex ML models into a human-readable form (Lundberg et al., 2020). XAI has the capability to make the inner workings of ML models and the patterns discovered in data transparent to provide additional insights for decision-makers (Meske et al., 2020). Consequently, the application of XAI for decision support to mitigate anchoring bias has already been proposed on a conceptual level (Wang et al., 2019).

What is currently missing in IS research within the context of anchoring bias, AI, and XAI trickles down to the following two aspects: First, the potential of AI decision support to aid individuals and its ability to overcome anchoring bias remains a widely under-investigated topic. Prior research has focused primarily on AI-based detection of the anchoring bias without explicitly supporting individuals in actual task accomplishment (Echterhoff et al., 2022). Hence, it remains unclear how tailored AI support displayed to users affects anchoring bias and, thus, individuals' task performance. Second, the ability of XAI to support individuals through the exposure of patterns that AI detects in data and its effect on anchoring bias. To date, the potential of XAI to provide such decision support has only been outlined conceptually (Wang et al., 2019) but not empirically tested with respect to anchoring bias.

Our study aims to shed light on this research gap by employing AI and XAI-based decision support configurations and two forms of anchors for a case where anchoring bias frequently occurs, namely e-commerce (Moser, 2020; Simonson & Drolet, 2004; Wansink et al., 1998). Relying on online experiments, we randomly assign participants to a decision support configuration (between-subject condition) and ask them to estimate a fair market price for six used car offerings of a German car online portal. Our AI and XAI-based decision support is intended to aid participants in evaluating the price for a given car offering. To measure the isolated and combined effect for a price and unrelated anchor, we conduct two distinct experiments, randomly employing specific anchor configurations for half of the six market estimation tasks (within-subject condition). The results of our experiments show that XAI helps to estimate a fair market price (i.e., increased task performance) when an offer does not contain an offer price (i.e., no price anchor). However, contrary to our expectations, sole XAI decision support (i.e., explanations only) can neither lower the anchoring bias from a price anchor nor from an unrelated anchor (set individually or jointly). Conversely, an AI price evaluation indicating a displayed price's fairness and its combination with XAI helps individuals estimate fair market prices despite the presence of anchors.

Our work contributes to the ongoing debate on the value of AI and XAI for human decision-making (Bauer et al., 2021; Berente et al., 2021). By providing initial insights on the effect of these technological advances on anchoring bias, we take a new perspective on strategies for overcoming cognitive biases. We expect our work to have concrete implications for the design of AI and XAI-based decision support and IS to mitigate anchoring bias to alter individuals' behavior.

# Background and Hypotheses

Starting from the theoretical background of anchoring bias and current developments in AI and XAI, we formulate hypotheses as a foundation for our experiment. Given the empirical evidence for anchoring bias in the purchasing context (Li et al., 2021) and the frequent exploitation of the phenomenon in practice, we choose the market price estimation of used cars as the experimental task of our study, to which the following hypotheses are related.

## Anchoring Bias

The anchoring bias is a phenomenon that refers to the influence of initially presented information (i.e., the anchor) on subsequent decisions or estimations (Tversky & Kahneman, 1974). Described as one of the most robust cognitive heuristics, the anchoring bias can be observed for novice as well as domain expert decision-makers (Northcraft & Neale, 1987). Different explanations have been given for the underlying psychological mechanisms of the anchoring bias (Furnham & Boo, 2011). The original study by Tversky and Kahneman (1974) used the term "anchoring and adjustment" for the bias, with the reasoning that the anchor is used as a starting point from which the decision maker adjusts towards a plausible value for answering the decision task. This adjustment process is effortful and typically insufficient; thus, the resulting answers are biased toward an initial anchor. Subsequent research established that the adjustment process is insufficient for explaining the anchoring effect and proposed the selective accessibility model (Strack & Mussweiler, 1997; Chapman & Johnson, 1999): decision-makers test the anchoring value as a hypothetical answer for the decision task and thereby construct a mental model and elicit information that is consistent with it. Even when the anchor itself is later discarded as a fitting answer, the mental model and related consistent information remain highly accessible for the following judgments and thus foster anchoring bias. Because the biasing information is generated by the decision makers themselves instead of external sources, they may fail to correct for the bias, which explains its robustness in a variety of environments (Strack & Mussweiler, 1997).

An area of research that has received particular attention is anchoring manipulation in an economic context, where the deliberate use of anchoring can have negative effects on consumer welfare. Examples for this are often found in e-commerce, e.g., using pre-discounted products as an anchor with the goal of triggering impulsive purchases (Moser, 2020), or extending willingness-to-pay and purchase quantity over initially held levels (Simonson & Drolet, 2004; Wansink et al., 1998). In online shopping platforms, it is especially easy to expose customers to price anchors, e.g., through advertisement banners (Wu & Cheng, 2011), which can act as a reference that affects customers' price beliefs (Biswas & Blair, 1991).

Thus, as a first step, we aim to replicate the findings of prior research in the context of online shopping by incorporating three forms of anchor configurations. First, the initial offering price represents a strong anchor for market valuation of buyers, as well as subsequent transaction prices (Bokhari & Geltner, 2011). As price negotiation is a traditional way of establishing a transaction price for used goods in general and the used car market in particular (Huang, 2020), we expected that sellers set high initial offers above a market price (i.e., an upward anchor). Thus, we hypothesize that an initial asking price (referred to as "upward price anchor" in the following) acts as an anchor that increases market price estimations:

*H1a: The presence of an upward price anchor induces anchoring bias and thus increases individuals' market price estimation.*

Second, even the use of unrelated anchors, e.g., prices from different categories of products (Adaval & Wyer, 2011), can lead to changes in price estimations. In online shopping portals, such unrelated anchors can be implemented, for example, as advertisements (Wu & Cheng, 2011). Li et al. (2021) conducted a meta-analysis in the purchasing context to explore the determinants of anchoring bias and found, among others, information relevance to be one of the most important factors for the biases' magnitude. Information relevance implies that an anchor conveys relevant information for the target response, i.e., a non-random anchor directly related to the target product (e.g., a price for price estimates) (Li et al., 2021). To cover multiple forms of anchors, we therefore additionally examine the effect of an unrelated but not contextually distant anchor that is not informationally relevant for the subjects' price estimates. The purchasing context is typically characterized by using anchors above market prices to manipulate customers' price beliefs and purchase decisions (Nunes & Boatwright, 2004). Thus, we expect a high unrelated upward anchor in the form of an ad from a different product category to elevate the value of participants' price estimation:

*H1b: The presence of an unrelated upward anchor induces anchoring bias and thus increases individuals' market price estimation.*

Third, we investigate the interaction of the unrelated anchor (advertisement) and the related and fully informationally relevant price anchor. Prior research posits that the existence of multiple anchors will reduce the reliance of the decision maker on any one anchor (Whyte & Sebenius, 1997). Additionally, multiple anchors may reduce the effect of unrelated anchors if both related and unrelated anchors are present (Whyte & Sebenius, 1997). Hence, we hypothesize the following:

*H1c: The presence of both an unrelated upward anchor and an upward price anchor reduces the anchoring bias of the unrelated anchor.*

### AI and XAI-based Decision Support

Knowing the challenges associated with human decision-making, companies have invested in related IS for general decision support (Arnott & Pervan, 2014), business intelligence (BI) (Shollo & Galliers, 2016), and, more recently, in ML-based AI applications to provide decision-aid (McKinney et al., 2020). IS employing such ML-based AI decision support (hereafter described as AI decision support) brought impressive examples of aid for management decisions in an organizational setting, e.g., in financial risk assessment (Wu et al., 2022) and demand planning (Gonçalves et al., 2021).

In addition to aiding organizations, prior research on human-AI collaboration has shown that AI decision support can increase task performance also on an individual level (Wilson & Daugherty, 2018). Examples range from rather simple tasks such as image classification (Fügener et al., 2021) to high-stake decisions such as breast cancer detection (Wang et al., 2016). Indeed, economists expect that this increase in task performance holds for many domains where individuals are provided with AI decision support (Gownder et al., 2017). In the context of market price estimation, AI has the capability to support individuals in providing targeted advice on the fairness of a displayed offer price. Hence, given that the presence of an AI decision support configuration seems to be positively correlated with task performance across various domains and tasks, we also expect an increase in individuals' performance in market price estimation:

*H2a: AI decision support leads to an increase in individuals' task performance in market price estimation.*

IS research has already shown that decisions can be supported in a way that helps to overcome cognitive biases (Küper et al., 2023). For example, a study on anchoring bias by Ni et al. (2019) indicates that decision support from a BI system effectively mitigates the effect of a spurious anchor (i.e., an unrelated anchor irrelevant to the task) but not for a task-related anchor. Given the potential of IS employing AI for human decision-making, it may also open new avenues for overcoming cognitive biases to foster decision-making in a more rational manner. So far, however, research on the mitigation of anchoring through AI decision support is limited to a few studies. One example is a study by Echterhoff et al. (2022) that introduced an approach for mitigation in sequential decision-making when individuals are biased from previous choices; they could show that the negative effect of anchoring bias can be decreased when tasks are presented in an order defined by AI.

For market price estimation, we assume that an AI-based price fairness evaluation has the capability to overcome the negative effect of anchoring bias. Lighthall et al. (2015) suggest emphasizing that specific results may also point to other hypotheses, which in turn encourages discarding one's own initial hypotheses (as those generated by anchors) and overwriting existing ones (Wang et al., 2019). Following this argumentation and the results from prior research on IS and AI to overcome anchoring bias, we assume that AI decision support also has the capability to overwrite the negative effect of positioned anchors by triggering rethinking processes. Also, through additional information provided by AI-based decision support, individuals may fully rely on the provided support and focus less on other unrelated anchors. Hence, we suggest the following hypothesis:

*H3a: AI decision support decreases the negative effect of anchoring bias in individuals' market price estimation.*

Although AI promises significant benefits for human decision-making, underlying ML models often remain opaque black boxes, leaving their power to support decisions through the exposure of captured patterns untapped (Meske et al., 2020). The stream of research that is concerned with the rationale behind AI decisions and their translation into a human-readable form goes by the notion of XAI (Barredo Arrieta et al., 2020). Research frequently defines XAI as a complementary tool to AI decisions, e.g., "XAI is a research field that aims to make AI systems results more understandable to humans" (Adadi & Berrada, 2018, p. 52139). However, some are also interested in exploring XAI as a decision support tool for exposing patterns in data, with a focus on providing support solely through explanations (i.e., providing explanations but no predictions, e.g., Carton et al., 2020). In recent years, XAI research has put forth a wide variety of explanation methods—one of the most widespread are feature attribution methods that quantify a feature's impact on a model outcome (Lundberg et al., 2020) and satisfy the majority of properties for human-friendly explanations (Molnar, 2019). Feature attribution explanations can support decision-makers in understanding the contribution of features to a specific ML-based prediction (e.g., Senoner et al., 2022).

While some studies document an intriguing increase in task performance for XAI-supported decision-making (Bansal et al., 2021; Lai et al., 2020), others found evidence for a decrease in performance (Alufaisan et al., 2021; Carton et al., 2020). In this regard, Wang & Yin (2021) note that domain expertise seems to play a crucial role in increasing task performance through XAI, especially when feature attribution methods are employed. Concerning this overall ambiguity of results, a recent meta-analysis by Schemmer et al. (2022) subsumes various studies across different domains and finds an overall positive effect of XAI-based support on human task performance.

Although we are aware that the value of XAI for decision-making is highly task-specific, we assume that support in the form of feature attributions can increase individuals' task performance by providing targeted guidance on which product characteristics (i.e., features) are particularly important for a market price estimate. Building on this assumption and recent literature on XAI-supported decision-making, we hypothesize:

*H2b: XAI decision support leads to an increase in individuals' task performance in market price estimation.*

Given XAI's ability to aid the decision of individuals through targeted explanations, there is a growing body of research investigating whether such explanations can also mitigate common heuristic biases (Bauer et al., 2021; Schemmer, Kuehl, et al., 2022). In this context, Wang et al. (2019) propose that feature attribution explanations might have the potential to overcome anchoring bias by avoiding confirmation and early closure (i.e., fixating on an initial decision). They assume that this is facilitated by encouraging individuals to explore alternative decisions as they are exposed to how different feature attributions are contrasted (Wang et al., 2019). This assumption may hold for the task of market price estimation, given that XAI-based feature attribution explanations can potentially overcome the negative effect of anchoring bias, as decision-makers reconsider initial decisions for an estimated price by focusing on important features of a specific product. Hence, we hypothesize the following:

*H3b: XAI decision support decreases the negative effect of anchoring bias in individuals' market price estimation.*

Building on the hypotheses above, we assume that AI, through price evaluations, and XAI, through explaining variables influencing a market price considered as appropriate, support mitigating the effect of anchoring bias and thus also increase task performance. Hence, we hypothesize that joint AI and XAI decision support will also have a desirable effect on market price estimations:

*H2c: Joint AI and XAI decision support leads to an increase in individuals' task performance in market price estimation.*

*H3c: Joint AI and XAI decision support decrease the negative effect of anchoring bias in individuals' market price estimation.*

### *Hypotheses Summary*

The overall research objectives of our research are to examine (i) the effect of AI and XAI-supported decision-making on individuals' task performance in the context of market price estimations and (ii) whether such support can decrease the negative effect of the anchoring bias. Building on these objectives, we summarize the research model in Figure 1.

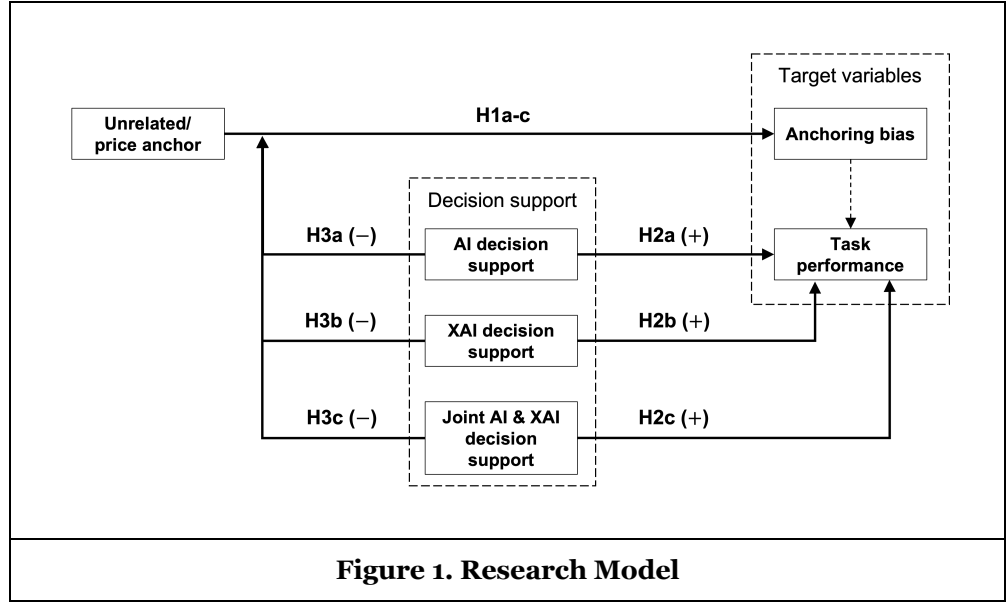

**Figure 1. Research Model**

We organize the hypotheses within our research model in three blocks: First, we assume that we observe an anchoring bias when anchors are set during individuals' market price estimation. To examine the effect of an anchor on the anchoring bias for our preceding hypotheses, our study employs an unrelated anchor, a price anchor, and both combined for various tasks (**H1a-c**). Second, we hypothesize that decision support from AI, XAI, and both combined can assist individuals in properly assessing displayed offer prices and thus positively affect individuals' ability to determine an adequate market price (i.e., task performance) (**H2a-c**). Finally, we assume that AI, XAI, and joint decision support can overwrite price and unrelated anchors, and thus reduce the negative effect of anchoring bias on decision-making (**H3a-c**).

Evaluating the effect of AI and XAI on human decision performance is a long-term goal of IS research (Bauer et al., 2021; Berente et al., 2021). So far, current studies neglect to evaluate the effect of AI and XAI on decision support in the context of heuristic biases and, especially, the anchoring bias. Hence, assessing the potential of these technological advances to mitigate the negative effect of anchoring bias is of particular interest to the IS field.

## Methodology

We conducted two online experiments in the context of used cars. In each experiment, all participants completed six tasks in which they estimated the actual market price of a displayed car offering. For this purpose, we displayed real online listings of used cars from a respective trading platform. For both experiments, we employ a mixed design in which we vary the type of anchor (no anchor, price anchor, and unrelated anchor) as a within-subject condition and the decision support provided as a between-subject condition (Figure 2). Our between-subject conditions comprise four configurations: first, no decision support (i.e., the control group); second, AI decision support (i.e., a traffic light that evaluates the fairness of an offer price); third, XAI decision support (i.e., a plot explaining the contribution of a car's feature to a price prediction); and fourth, combined AI and XAI decision support. Although we are aware that XAI is often understood as a complementary tool to AI (Adadi & Berrada, 2018), we are also, following Carton et al. (2020), interested in investigating the marginal effect of feature attribution explanations on task performance and anchoring bias. Hence, we introduce XAI as a separate decision support configuration.

| | | Within-subject condition | | | |
|---|---|---|---|---|---|
| | | **Experiment 1** | | **Experiment 2** | |
| | | Price & unrelated anchor | Price anchor | Unrelated anchor | No anchor |
| **Between-subject condition** | **No decision support** | 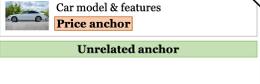 1-1 Car model & features / Price anchor / Unrelated anchor | 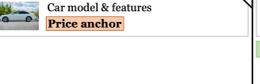 1-5 Car model & features / Price anchor | 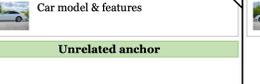 2-1 Car model & features / Unrelated anchor | 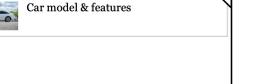 2-3 Car model & features |
| | **AI decision support** | 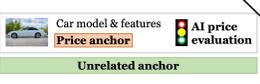 1-2 Car model & features / Price anchor / AI price evaluation / Unrelated anchor | 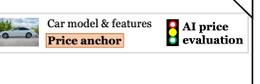 1-6 Car model & features / Price anchor / AI price evaluation | | |
| | **XAI decision support** | 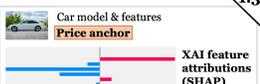 1-3 Car model & features / Price anchor / XAI feature attributions (SHAP) / Unrelated anchor | 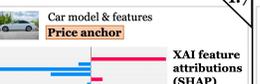 1-7 Car model & features / Price anchor / XAI feature attributions (SHAP) | 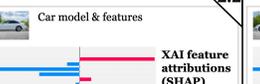 2-2 Car model & features / XAI feature attributions (SHAP) / Unrelated anchor | 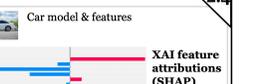 2-4 Car model & features / XAI feature attributions (SHAP) |
| | **AI & XAI decision support** | 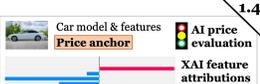 1-4 Car model & features / Price anchor / AI price evaluation / XAI feature attributions (SHAP) / Unrelated anchor | 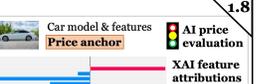 1-8 Car model & features / Price anchor / AI price evaluation / XAI feature attributions (SHAP) | | |

**Figure 2. Experimental Conditions of Experiment 1 and 2**

We separated our study into two experiments to avoid contamination effects of the price anchors across the six estimation tasks, as the price anchor of previous task could have acted also as an anchor for a subsequent task without price anchors.

Our *first experiment* examines anchoring bias by measuring the impact of a sole price anchor and both an unrelated and price anchor on participants' price estimations (**H1a, H1c**). Within this experiment, we also explore the general potential of AI, XAI to and the joint decision support to improve task performance (**H2a-H2c**) and to mitigate the negative effect of anchoring bias (**H3a-H3c**). We use a 2 x 4 full factorial design in which participants completed tasks with either a price anchor or a price and an unrelated anchor (within-subject condition) while keeping the decision support per participant constant (between-subject condition). Our first experiment employs all four decision support configurations: no decision support, AI decision support (i.e., the traffic light), XAI decision support consisting of a SHAP-based feature attribution explanations (Lundberg et al., 2020), and a joint AI and XAI decision support configuration. This splits the participants into four experimental groups to which participants were randomly allocated.

Our *second experiment* omits the price anchors to examine the effect of the unrelated anchor (**H1b**) and XAI decision support in more detail (**H2b, H3b**). We employ a 2 x 2 experimental design in which participants completed their tasks with either no anchor or an unrelated anchor (within-subject condition) and two different decision support configurations (between-subject condition): no decision support and XAI decision support. As there are no price anchors in the second experiment, the AI decision support (which evaluates the fairness of a displayed price) does not contain relevant information and is therefore omitted.

### Experiment Implementation and Stimuli

We implemented the experiments as an online survey. Participants first entered their demographic data, answered questions on their domain literacy, and received an introduction to the tasks as well as a short explanation for the respective decision support. Thereafter, we presented the market price estimation tasks to the participants.

## Experimental Tasks

We explain the experimental task using the joint AI and XAI decision support condition and both the price and unrelated anchor in Figure 3 using an exemplary task containing all the stimuli of our study.

**Figure 3. Exemplary Task (Translated From German Language)**

We displayed six *real listings of used cars (1)* to all participants in a random order. Our listings were obtained from the trading platform *Autoscout24.de*. At the end of each task, the study participants were asked to *estimate the actual market price for the given offer (2)*. We selected the most frequently registered car models in the segments "small cars" and "executive cars" of the German Federal Motor Transport Authority (2023). As a result, we used three listings each for the models Opel Corsa (segment small cars) and Audi A6 (segment executive cars).

We employed two types of anchors: First, a *price anchor (3)*, which originates from the real car listing prices (Table 1). Second, an *unrelated anchor (4)* within an advertisement banner below the car details. We thereby stayed in the context of cars and used a fictitious advertisement for garages linked to an actual seller website to make the advertisement look natural. We chose a high anchor value and embedded a 60,000€ price tag within the advertisement. Following the definition of anchoring as the initial piece of information presented (Tversky & Kahneman, 1974), we displayed a loading spinner for the *used car listing (1)* in the first three seconds of each task so that the participants were initially exposed to the unrelated anchor.

Our between-subject conditions rely on an *AI (5)* and an *XAI decision support (6)*. The AI and XAI decision support is based on ML models that predict an actual fair market price for a used car based on its features. For model training, we use a publicly available dataset from Autoscout24.de, which contains 34232 listings of used cars in Germany for the year 2021. To provide a precise market price estimation, we applied multiple ML algorithms, namely CatBoost (Prokhorenkova et al., 2018), XGBoost (Chen & Guestrin, 2016), Random Forest (Breiman, 2001), and compare them against linear regression as baseline estimator. We divide our entire dataset into a train-test dataset (80%) to determine the best parameters of the algorithms using a 5-fold grid search cross-validation. Thereafter, we rely on a dedicated validation dataset (the remaining 20% of the data) to select the best ML approach. Among the ML algorithms, CatBoost yields the highest predictive performance[1], resulting in significantly lower absolute residuals than linear regression ($t(6845)$ = 21.53, $p$ < .001, $d$ = 0.368). Using CatBoost, we train a model for each of the two car segments, respectively. We do so as our selected car models represent two very distinct market segments and thus represent different price ranges. Hence, the AI-based price evaluations and feature attributions (XAI decision support) are more meaningful when considered separately for each price segment.

The AI decision support consists of a price fairness evaluation that relies on the difference between an offered price (i.e., listing price) and the predicted market price considered fair. The listing price evaluation is embodied *within a traffic light (5)* that indicates the price to be unfair (red), moderate (yellow), or fair (green). An offer price with a difference of equal or less than 10% of the predicted price is evaluated as fair, with a difference equal or lower than 20% as moderate, and with a difference higher than 20% as unfair (Table 1). For the XAI decision support, we applied the XAI method SHAP (Lundberg et al., 2020) and generated force plots to *explain the price prediction based on a car's features (6)* (Lundberg et al., 2020). The SHAP force plot visualizes the feature attributions for the car's price prediction while having the segment's average predicted price as a reference line in the middle (Figure 3).

| Task | Market segment | Car model | Listing price [Euro] | Predicted price considered fair [Euro] | Price fairness |
|------|----------------|-----------|----------------------|----------------------------------------|----------------|
| 1 | Small cars | Opel Corsa | 7500 | 5193 | unfair |
| 2 | Small cars | Opel Corsa | 5990 | 5220 | moderate |
| 3 | Small cars | Opel Corsa | 8480 | 8256 | fair |
| 4 | Executive cars | Audi A6 | 22690 | 16652 | unfair |
| 5 | Executive cars | Audi A6 | 28900 | 25502 | moderate |
| 6 | Executive cars | Audi A6 | 24999 | 22981 | fair |
| **Table 1. Experimental Tasks** | | | | | |

## Sample Description and Experiment Variables

We recruited 260 participants from consumerfieldwork.de for experiment 1 and 130 from prolific.co for experiment 2 to avoid sample overlap, which adds up to N=390 study participants. As a constraint for our study, we set the minimum age for participants to 18 years and controlled for the representativeness of the German population regarding gender, age, and educational degree.

In summary, our sample of both experiments is balanced regarding gender (48.5% female, 51.3% male, 0.3% diverse), which is similar to the gender distribution of the German population. The distribution of age categories (11.8% 18-24 year-olds, 38.2% 25-39 year-olds, 25.4% 40-59 year-olds, and 24.6% over 60-year-olds) is slightly biased towards younger participants in our study: Our sample contains 50% participants between 18 and 39 years, while these two younger age categories account for only 30% of the German population of over 18 year-olds (DESTATIS, 2022). Regarding the educational degree, our sample comprises 50.8% of participants with a higher degree, 24.9% with a middle degree, and 24.4% with a lower degree. This is fairly representative for the German population, although slightly shifted towards participants with a higher degree (33% in the German population hold a higher degree; DESTATIS, 2023).

For both experiments, we additionally collected variables that may have a major impact on our target variables. Thereby, we focus on participants' knowledge in the car domain, given that previous studies have shown that existing pre-knowledge, expertise, and experience have a significant impact on task performance (Wang & Yin, 2021) and human susceptibility to anchoring bias (Furnham & Boo, 2011; Wilson et al., 1996). For domain literacy, we formulated questions on participants' knowledge and interest in cars, the automotive market, and their general knowledge on car prices. For each question, we included a 5-point Likert scale from 1 – very low to 5 – very high. Furthermore, we asked participants about their amount of purchased and sold cars, and if they deal with the car domain professionally.

---

[1] Results on the validation dataset: Mean Absolute Percentage Error (MAPE) = 10.012%; Mean Absolute Error (MAE) = 1200.50 (total price range: 1100€ - 167300€).

# Results

To increase the data quality and internal validity of our study, we exclude 19 out of 390 participants from our analyses because they provided unrealistically low estimates for car prices (below €500).

## Hypotheses 1a-c

To test our hypotheses on anchoring bias in the given experimental setting, we conduct two-sample one-sided *Welch's* t-tests and report the effect sizes using *Cohen's d* (Table 2). For the analysis of price anchors (H1a), we compare individuals' market price estimations in experiment 1 (with price offers that act as an anchor) to estimations in experiment 2 (no price anchor displayed). We consider only answers of the experimental groups that are present in both experiments (XAI and no decision support) that did not receive an unrelated anchor; hence, we compare the groups 1.5 and 1.7 with the groups 2.3 and 2.4 (see Figure 2). For the analysis of unrelated anchors (H1b), we compare all estimates made with unrelated anchors to those that were not affected by an unrelated anchor. To avoid contamination with price anchors, only estimates in experiment 2 were considered; therefore, we compare the groups 2.1 and 2.2 with the groups 2.3 and 2.4. To test the third hypothesis (H1c), we examine the combined influence of a price anchor and an unrelated anchor to estimations made with a price anchor only (i.e., comparing the groups 1.1-1.4 with the groups 1.5-1.8).

| | Sample 1 | Sample 2 | p-value | Eff. size (Cohen's d) |
|---|---|---|---|---|
| *H1a* | Groups = 1.5 and 1.7 Mean (SD) = 10925.18 (8032.03) | Groups = 2.3 and 2.4 Mean (SD) = 9418.83 (6984.43) | 0.003 | 0.197 |
| *H1b* | Groups = 2.1 and 2.2 Mean (SD) = 10419.1 (8888.29) | Groups = 2.3 and 2.4 Mean (SD) = 9418.83 (6984.43) | 0.045 | 0.125 |
| *H1c* | Groups = 1.1 − 1.4 Mean (SD) = 11256.01 (8313.30) | Groups = 1.5 − 1.8 Mean (SD) = 11384.39 (9069.71) | 0.612 | 0.015 |
| **Table 2. Differences in the Mean of the Treatment Conditions and the Control Group** | | | | |

We find evidence for the hypotheses H1a and H1b as both the unrelated anchor and the price anchor have a statistically significant upward impact on the price estimates. The resulting small effect sizes of our study fall in line with the results of a meta-analysis on anchoring bias related to willingness-to-pay while being lower than the reported mean of 0.27 (Li et al., 2021). If price anchors and unrelated anchors are combined, the price estimates do not differ significantly from estimates made with price anchors only; therefore, the unrelated anchors no longer show an effect on price estimates, and H1c is supported. Consequently, we do not analyze the effect of decision support on unrelated anchors in experiment 1 in later analyses.

## Hypotheses 2a-c

We measure task performance as the absolute difference between the estimated price and the actual market price determined by our ML model. To formally test the impact of the decision support conditions on task performance, we subsequently estimate one ordinary least squares model per experiment, with the control group (no decision support) as a reference. Thereby, we control for domain literacy and the six individual estimation tasks (Table 3).

We find that price estimates were generally more accurate in experiment 1 (containing price anchors) compared to experiment 2 (no price information), as indicated by the lower absolute estimation error in the control group. In the presence of price anchors of experiment 1, the participants assigned to decision support conditions (AI, XAI, and AI and XAI) display a lower error in their estimations compared to the control group; however, the differences are not statistically significant. In experiment 2 without price anchors, XAI decision support leads to less error in the price estimation and, thus, higher task performance ($p = .059$). In both experiments, higher domain literacy is associated with a more accurate price estimation. In summary, we do not find evidence to support hypotheses H2a and H2c and find partial support for hypothesis H2b.

| Independent variable | Absolute estimation error [Euro] | |
|---|---|---|
| | Experiment 1 | Experiment 2 |
| Intercept (Reference: No support) | 2832.64*** (506.63) | 4419.99*** (679.83) |
| Treatment AI | -324.60 (349.24) | – |
| Treatment XAI | -344.75 (359.20) | -663.73* (350.63) |
| Treatment AI & XAI | -474.61 (365.24) | – |
| Domain Literacy | -316.44** (130.73) | -751.31*** (215.08) |
| R² | 0.196 | 0.438 |
| Observations | 1488 | 738 |
| **Table 3. Analysis of Decision Support on Task Performance** | | |

Notes: The table displays the effects of the decision support conditions on task performance (measured as absolute deviation from market price in euro), controlling for domain literacy and individual estimation tasks 1-6 (dummy variables not reported). Standard errors are presented in parentheses. *, **, and *** indicate significance at the 10%, 5%, and 1% level, respectively.

## Hypotheses 3a-c

For the analysis of the effect of our decision support on the anchoring bias, we estimate an ordinary least squares model with the deviation of participants' market price estimation to a price anchor as the dependent variable. We use the four decision support conditions as independent variables with the control group (i.e., no decision support) as a reference. The statistical model also estimates the impact of the three fairness levels of the tasks ("unfair", "moderate", "fair", with "fair" as the reference group) to capture the effect depending on the "traffic light" indicator of the AI decision support, as well as their interactions with the decision support conditions (Table 4).

| Independent variable | Difference of market price estimation to price anchor [Euro] | |
|---|---|---|
| | Coefficient | Standard Error |
| Intercept (Reference: No support) | 5465.16*** | 551.98 |
| Treatment AI | -1916.52** | 826.48 |
| Treatment XAI | -714.01 | 850.65 |
| Treatment AI & XAI | -2658.95*** | 864.48 |
| Task moderate | 1123.74 | 780.61 |
| Task unfair | -139.86 | 780.61 |
| Treatment AI × Task moderate | 406.49 | 1168.82 |
| Treatment XAI × Task moderate | 396.31 | 1203.00 |
| Treatment AI & XAI × Task moderate | 1192.41 | 1222.56 |
| Treatment AI × Task unfair | 2375.67** | 1168.82 |
| Treatment XAI × Task unfair | -296.78 | 1203.00 |
| Treatment AI & XAI × Task unfair | 2772.52** | 1222.56 |
| R² | 0.217 | |
| Observations | 1488 | |
| **Table 4. Analysis of Decision Support on Anchoring Bias** | | |

*Note: The table displays the effects of the decision support conditions on anchoring bias (measured as deviation from displayed price anchor in euro). *, **, and *** indicate significance at the 10%, 5%, and 1% level, respectively.*

We find that price estimates in the experimental conditions containing an AI decision support (i.e., a traffic light evaluating the offer price) differ significantly less from the value of the price anchor in the fair condition (i.e., a green traffic light). Participants in the AI decision support and AI and XAI decision support groups estimate prices that are 1916.52€ (35%) and 2658.95€ (49%) closer to the anchor compared to the control group. This stronger anchoring bias, however, depends on the level of task fairness: For tasks considered unfair (i.e., a red traffic light), the results display a significantly diminished anchoring effect, whereas AI decision support in the tasks with moderate fairness ("yellow light") shows no statistically significant difference compared to fair tasks. Consequently, the AI decision support and its combination with XAI works as intended (i.e., mitigates anchoring for unfair tasks and thus the negative effect of anchoring bias), which supports H3a and H3c.

For market price estimates made with XAI decision support, we find no statistically significant changes in price anchoring or interactions with task fairness and, thus, no support for H3b. This finding is further reinforced when investigating the effect of XAI on unrelated anchors from experiment 2: A two-sample *Welch's* t-test comparing estimates with no decision support (group 2.1) and XAI decision support (group 2.2) reveals no significant differences in market price estimates ($p$ = .851).

## Discussion

Our experimental evaluation led to three main findings outlined in Table 5. Within this section, we discuss these findings, name limitations, and formulate implications as well as future research needs for the field of AI and XAI-based decision support and IS to overcome anchoring bias.

| Main findings | Implications for the design of AI/XAI-based decision support | Implications for IS to overcome anchoring bias |
|---|---|---|
| XAI decision support does increase task performance when no price (i.e., no price anchor) is displayed | XAI decision support can provide valuable insights in a decision environment when no distorting information is set (e.g., a price) | |
| Sole XAI decision support does not reduce the effect of anchors on individuals' market price estimates | | Sole XAI decision support is not effective in mitigating the negative effect of anchoring bias |
| Sole AI decision support and its combination with XAI do reduce the effect of an unfair price anchor on individuals' market price estimates | Targeted AI decision support seems to be effective in evaluating prices (i.e., price anchors)<br><br>XAI decision support must be contextualized (i.e., be related to AI support) to be effective | AI decision support and its combination with XAI decision support can be effective in mitigating the negative effect of anchoring bias |
| **Table 5. Overview of Findings and Implications From Our Study** | | |

### *Summary of Main Findings*

Recently, there has been a great interest in the technological advances of AI and XAI to support individuals' decision-making (Bauer et al., 2021; Berente et al., 2021). While IS research has already put forth tangible examples for such support (e.g., Fügener et al., 2021), it remains widely unclear how these technological advances can help to avoid common cognitive biases such as anchoring bias. Hence, prior studies have not yet empirically evaluated the potential of AI and XAI-based decision support to overcome anchoring bias (Wang et al., 2019).

Our study addressed this research gap by conducting two experiments in the context of e-commerce. We asked a total of N=390 participants to estimate the market price of six used cars and assigned participants to an AI, XAI, or joint decision configuration, while manipulating decisions through a price and unrelated anchor for half of the six tasks. Overall, we identify three main findings from our study, leading to several implications for research on AI and XAI-based decision support and IS to mitigate anchoring bias.

Our first main finding is that *XAI decision support does increase task performance when no price (i.e., no price anchor) is displayed*. In experiment 2, we compared XAI decision support with a group that received no decision support and found an increase in task performance; however, we observed a contrary result in experiment 1. We attribute this to the circumstance that in experiment 2, the decision environment might be more artificial, i.e., it does not reflect a realistic scenario for buying a used car on an online marketplace since the tasks did not contain offer prices. This might account for a stronger isolated effect of XAI decision support on individual task performance, as the tasks do not include a reference point (i.e., a price anchor). As the anchoring bias describes the tendency to rely heavily on a single piece of information (Kahneman, 2011), such explanations might only be effective in the absence of the rather strong effect of an (unfair) price anchor (see analysis on H1a). Hence, we find that for research on the design of such decision support, XAI can provide valuable insights to individuals. However, one might need to account for the salience of such explanations as distorting information seem to have a non-negligible impact on individuals' task performance.

The second main finding is that *sole XAI decision support does not reduce the effect of anchors on individuals' market price estimates*. Thereby, we found in experiment 1 that XAI decision support had no mitigating effect when a task included a price and an unrelated anchor. Likewise, in experiment 2, where we examined the isolated effect of an unrelated anchor, we did not find any XAI decision support mitigation capabilities. This indicates that, so far, sole XAI decision support is not effective in mitigating the negative effect of anchoring bias. This contrasts with the assumption of Wang et al. (2019) that feature attribution explanations encourage individuals to explore alternative decisions as they learn how different features contrast with one another, which might reduce anchoring bias—however, one might need to consider that Wang et al. (2019) relate their XAI support to the output of an ML-based diagnostic tool (i.e., joint AI and XAI decision support). We conclude that for our case and a sole XAI decision support, this assumption does not hold.

Finally, our third main finding is that *sole AI decision support and its combination with XAI reduce the effect of an unfair price anchor on individuals' market price estimates*. We observe this result through the analyses of experiment 1, which shows that through AI support, individuals recognize unfair prices. Also, the traffic light for price evaluation combined with XAI support seems to reinforce the reducing effect of a price anchor on individuals' market price estimates. These results have two implications for the design of AI and XAI-based decisions: First, the AI decision support may be effective in evaluating prices because there is a direct fit between the task and the implemented support (i.e., the traffic light). One of the existing explanations in IS research for individual performance improvement through AI is the task-technology fit theory introduced by Goodhue & Thompson (1995), which states that technology and its use must fit well with the tasks it supports to affect individuals' performance positively (Sturm & Peters, 2020). Second, building on the previous implication, XAI can enhance AI's abilities to mitigate anchoring bias when put in the context of the support provided by AI. In general, we conclude that AI and its combination with XAI are effective for reducing the negative effect of anchoring bias.

### Limitations and Future Work

Despite our best efforts, our study has several limitations. First and foremost, one potential weakness is the generalizability of our results beyond the case of market price estimations on used car marketplaces. Given that the anchoring bias also appears in various domains (Furnham & Boo, 2011), the effects of AI and XAI decision support might be context and case-specific. Participants may generally have the attitude that the asking prices on used car marketplaces are set too high (i.e., as sellers could strategically use the offer price as an anchor to steer the negotiation process to their advantage), which can also influence price estimates. In addition, participants might have been more likely to adjust their price estimates downwards when the AI decision support suggests an "unfair" or "moderate" price. Therefore, our results might not be generalizable to use cases such as electric load forecasting, where adjustments of estimates in both directions (i.e., upwards and downwards) are more meaningful and likely (Giacomazzi et al., 2023). Although we paid attention to obtain a sample that is representative of the German population, our sample was slightly biased towards younger and more educated participants, which might further hamper the generalizability of our results.

Further limitations concern the implementation of technical aspects: Our XAI component relies only on default SHAP force plots, which other research has found difficult to interpret (Wastensteiner et al., 2021), potentially leading to limited value for decision-making. Thus, more task-specific visualizations for price estimations (e.g., including the value of feature attributions as monetary units) and other XAI visualization methods, such as LIME (Ribeiro et al., 2016), could be implemented to improve the decision support.

A final concern is the lack of moderating variables in our research model. Although we collected the domain literacy of all participants (which significantly influenced task performance), additional variables could provide further insight into the conditions under which AI and XAI successfully interact with anchoring and task performance. Such variables could be self-reported or measured attention towards the decision support components, perceived usefulness, perceived price fairness, or the attitude towards AI.

Future work could address the limitations of our study by adapting our research design, for example, to test more use cases in order to improve the generalizability of our results. Additionally, participants could receive feedback on their task performance after each task, which might further improve their decision-making. Extensions of our work may also alter the XAI decision support through more tailored support (e.g., by testing different use case-specific visualizations), consider collecting more variables to analyze them as moderators (e.g., AI attitude), and more thoroughly investigate the psychological mechanisms behind anchoring bias in IS. Another interesting aspect in the context of our study would be to explore the extent to which XAI and AI itself generate anchors and thus influence human decision-making.

## Conclusion

Recent advances in AI and XAI have opened opportunities for mitigating cognitive biases (Berente et al., 2021; Schemmer et al., 2022). However, the investigation of such technologies with respect to anchoring bias, one of the most robust and prevalent heuristics, has only been outlined conceptually (Wang et al., 2019) but not empirically tested. In our paper, we contribute to this stream of research by investigating the effect of AI and XAI decision support on the anchoring bias. To this end, we implement such support for a market price estimation task and show which support configuration can mitigate the negative impact of positioned anchors in an online experiment with 390 participants. Our results suggest that XAI decision support does increase task performance when no price (i.e., no price anchor) is displayed. However, sole XAI decision support does not reduce the effect of anchors on individuals' market price estimations. Regarding the mitigation of the anchoring bias, we find that AI decision support in the form of a traffic light reduces the effect of an unfair price anchor on individuals' market price estimates, either as the sole decision support condition or in combination with XAI decision support. Our findings have implications for the design of IS employing AI and XAI-based decision support that aim to overcome the anchoring bias. Although there are still many open questions that should be addressed in future research, we conclude that such decision support has the potential to improve task performance and, thus, mitigate the negative effect of anchoring bias.